\documentclass[11pt]{article}

\usepackage{epsfig}
\usepackage{citesort}

\newcommand{\vecpd}{$\vec{p} \vec{d}$}
\newcommand{\vecpp}{\( \vec{p} \vec{p} \)}

\newcommand{\ayp}{$A_y$}
\newcommand{\itll}{$iT_{11}$}
\newcommand{\cyy}{$C_{y,y}$}
\newcommand{\thcm}{\theta_{\mathrm{cm}}}

\begin{document}

\begin{center}
\Large
\textbf{Evidence for a three-nucleon-force effect in proton-deuteron
elastic scattering}

\vspace{0.2in}
\normalsize

%\author{
R.~V.~Cadman$^1$,
J.~Brack$^2$,
W.~J.~Cummings$^3$,
J.~A.~Fedchak$^3$,
B.~D.~Fox$^2$,
H.~Gao$^4$,
W.~Gl\"{o}ckle$^5$,
C.~Grosshauser$^6$,
R.~J.~Holt$^1$,
C.~E.~Jones$^7$,
E.~R.~Kinney$^2$,
M.~A.~Miller$^1$,
W.~Nagengast$^6$,
B.~R.~Owen$^1$,
K.~Rith$^6$,
F.~Schmidt$^6$,
E.~C.~Schulte$^1$,
J.~Sowinski$^8$,
F.~Sperisen$^8$,
E.~L.~Thorsland$^1$,
R.~Tobey$^1$,
J.~Wilbert$^6$,
H.~Wita\l a$^9$
%}

\vspace{0.2in}
\textit{
%\address{
$^1$Department of Physics, University of Illinois, Urbana, 
Illinois 61801\\
$^2$Nuclear Physics Laboratory, University of Colorado, Boulder, 
Colorado 80309-0446\\
$^3$Physics Division, Argonne National Laboratory, Argonne, 
Illinois 60439\\
$^4$Laboratory for Nuclear Science, Massachusetts Institute of Technology, 
Cambridge, Massachusetts 02139\\
$^5$Institut f\"ur theoretische Physik II, Ruhr-Universit\"at Bochum,
D-44780 Bochum, Germany\\
$^6$Physikalisches Institut, Universit\"at Erlangen-N\"urnberg, 
D-91058 Erlangen, Germany\\
$^7$W.K. Kellogg Radiation Lab, California Institute of Technology, 
Pasadena, California 91125\\
$^8$Indiana University Cyclotron Facility, Bloomington, Indiana
47408\\
$^9$Institute of Physics, Jagellonian University, PL-30059 Cracow,
Poland
}
\end{center}

\begin{abstract}
Developments in spin-polarized internal targets for storage rings have
permitted measurements of 197 MeV polarized protons scattering
from vector polarized deuterons.  This work presents measurements of
the polarization observables \ayp , \itll , and \cyy\  in 
proton-deuteron elastic scattering.  When
compared to calculations with and without three-nucleon forces, the
measurements indicate that three-nucleon forces make a significant
contribution to the observables.  This work indicates that three-body forces
derived from static nuclear properties appear to be crucial to the 
description of dynamical properties.
\end{abstract}
%\pacs{21.30.Fe,21.45.+v,24.70.+s,25.40.Cm,25.45.De}
% notes:
% 21.30.Fe: nuclear forces in hadronic systems and effective interactions
% 13.75.Cs: nucleon-nucleon interactions, but not in nuclei
% 21.45.+v: few body systems
% 24.70.+s: polarization phenomena
% 25.40.Cm: elastic proton scattering
% 25.45.De: elastic and inelastic deuteron scattering

% body of paper here
 Understanding how nuclei are built from their constituent protons and
neutrons and the forces between them is one of the fundamental goals of
nuclear physics.  An important aspect of the nuclear force is the
modification which occurs as nucleons become embedded in the nuclear
medium. The three-nucleon system provides an important laboratory because
the Schr\"{o}dinger equation, in the form of Faddeev equations, can be
solved exactly for three bodies.  New computational capabilities have
extended the range of validity for Faddeev calculations %\cite{faddeev}
to a wide range of kinematics, from bound states to scattering and 
three-body breakup at energies up to 200 MeV.

Modern two-nucleon potentials based on the exchange of pions and heavier
mesons \cite{Bonn-orig,av18-orig,nijmegen} provide an excellent
description of
neutron-proton and proton-proton scattering data and of
the deuteron.  In contrast, it is now generally accepted that the
binding energies of other light nuclei cannot be calculated from the
modern two-nucleon potentials alone
\cite{Bonn-orig,Friar:triton,Glockle:triton,helium4}.  The only
successful resolution of this problem has been to include additional
potential terms which act only in the presence of at least three nucleons
\cite{tucson,other3NF}. These terms, known as three-nucleon forces (3NF),
are also essential for understanding nuclear matter in extreme conditions,
such as dense nuclear matter and neutron stars \cite{neutronstar}.

Three-body forces are expected in nuclear physics because conventional
nuclear theory is a simplification of the fundamental theory of the strong
interaction, quantum chromodynamics.  
In conventional nuclear theory, the nucleons are
treated as fundamental particles, and the nucleon excited states, which
are an expression of the underlying quark degrees of freedom, are not
explicitly included.  The addition of nucleon excited states leads to
forces which cannot be reduced to successive two-nucleon interactions.
For example, a pion exchange between two nucleons can excite one nucleon
two a $\Delta$ which subsequently decays by pion exchange with a third
nucleon.  That process
was included in the first three-nucleon potential, proposed by Fujita and
Miyazawa in 1957 \cite{FM3NF}.  Modern three-nucleon potentials are more
extensive, and include terms which follow from a fundamental
symmetry of the strong interaction, chiral symmetry \cite{chiral}.

Although the binding energies imply that the 3NF is significant, they
only constrain its overall strength.  An investigation
of nucleon-deuteron ($Nd$) scattering is needed to study the dynamical
characteristics of the three-body force.  For example, since the momentum
transferred to the deuteron can be varied by changing the incident nucleon
energy and the scattering angle, it is possible to probe the spatial
dependence of
the 3NF by using this reaction.  Recent calculations have shown that the
Tucson-Melbourne (TM) 3NF
\cite{tucson} predicts a significantly enhanced differential cross
section in
$Nd$ elastic scattering when the kinetic energy of the incident nucleon
is greater than about 60 MeV \cite{smokingGun}. 

Previous work compared measurements of the proton analyzing power \ayp\ 
in \vecpd\ elastic scattering to Faddeev
calculations \cite{stephenson}.  These authors showed that for a deuteron
recoil angle $\theta_{\mathrm{lab}} = 42.6^\circ$ and over a proton energy
range from 120 to 200~MeV, the Tucson-Melbourne potential over-corrects
the prediction of the CD-Bonn potential.  They also showed earlier
measurements of \ayp\ at 200~MeV \cite{ab72,iucf93}, which demonstrate that
neither theory predicts the correct angular dependence for \ayp .
Another recent measurement showed the same result at 150 MeV \cite{riken}.
This indicates that the \ayp\ puzzle \cite{aypuzzle} first observed at
low energies \cite{pd-lowenergy} persists at higher energies.  The
discrepancy in \ayp\ at low energies is believed to be due in part to
an uncertainty in the $^3P_J$ $np$ phase shifts \cite{tornow} 
or to the need for a
tensor component in the three-nucleon force \cite{tensor3nf}.
Measurements of the deuteron analyzing power 
\itll\ at the equivalent of a proton energy
of 135 MeV \cite{riken} are in good agreement with the predictions
of CD-Bonn+TM3NF.  However, at that energy none of the predictions 
match the angular dependence of the tensor analyzing powers $A_{xx}$
and $A_{yy}$.
At 200~MeV, the previous measurement of
\itll\ \cite{garcon} is not sensitive enough to distinguish the effect
of the 3NF.  Preliminary measurements of the spin correlation parameter
\cyy\ at 200~MeV \cite{meyer} were made at small angles 
($\thcm < 60^\circ$) where the Tucson-Melbourne 
potential does not contribute significantly to the observables.  

In an effort to provide further constraints on the nuclear three-body
force, we provide measurements of the three spin observables
\ayp ,  \itll , and \cyy\ at a proton energy of 197 MeV and throughout the
angular range \( 65^\circ < \thcm < 115^\circ \), where the 3NF contribution
is expected to be large.
\cyy\  is a measure of the asymmetry between
beam and target spins parallel and anti-parallel, \itll\ is a 
measure of the asymmetry
associated with changing the sign of the deuteron's vector polarization,
and \ayp\ is a measure of the asymmetry
associated with changing the sign of the proton polarization.
The present work made use of the polarized proton beam at the Indiana
University Cyclotron Facility (IUCF) with
a vector polarized
deuteron target.  Both the beam and target polarizations were parallel 
to the $y$ axis or
perpendicular to the scattering plane.  Both polarizations were changed
regularly and independently, so that
the three different
polarization observables could be measured simultaneously.

The proton beam used in this experiment was accelerated to 197 MeV by
IUCF's two
cyclotrons and then stored in an electron-cooled proton storage ring. The
average beam current was about 75 $\mu$A.  The beam spin direction
was alternated between parallel and
anti-parallel to $\mathbf{\hat{y}}$ each time the ring was filled.

\begin{figure}[bt]
\centering{\epsfig{file=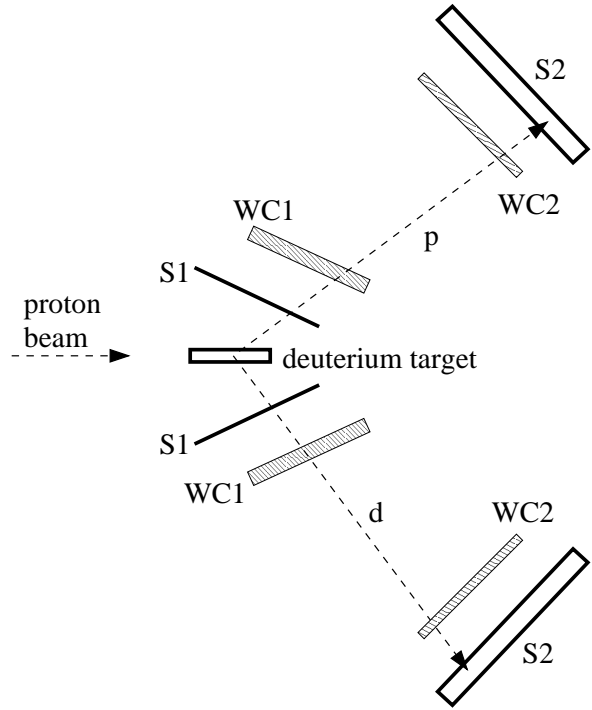,width=3.05in}}
\caption{The experimental apparatus, viewed from above.  The detectors are %
arranged symmetrically about the beam line.  Each side has a %
forward thin scintillator (S1), two \mbox{delay-line} wire chambers  %
(WC1 and WC2), and a stack of three thick scintillators (S2).  Possible %
trajectories of an outgoing proton (p) and deuteron (d) are also %
shown.  Not shown are the polarized source above the target %
cell and the atomic polarimeter below.} \label{fig:detectors}
\end{figure}

The experimental apparatus is shown schematically in Figure
\ref{fig:detectors}. The
detectors were symmetric about the beam line, and included a forward thin
scintillator (0.31 cm thick), two delay-line wire chambers, and a stack of
three 100 cm $\times$ 15 cm $\times$ 10 cm scintillator bars.  The wire
chambers consisted of two planes of wires, with wires running horizontal in
one plane and vertical in the other.  The wire spacing was 0.8 cm, which
led to an angular resolution of about 15 mr.  The momentum of the
scattered protons and deuterons was measured from the time of flight
between the thin scintillator and the thick scintillator array.  
Deuterons were distinguished from protons by the relationship between the
time of flight and the energy deposited in the thick scintillators.

Polarized deuterium atoms were
injected into the center of the target, a 40 cm $\times$ 3.2 cm
$\times$ 1.3 cm rectangular aluminum tube open at both ends to
allow the proton beam to pass.  The polarization of the target
was continuously monitored using a Rabi polarimeter
which sampled atoms from the center of the target cell.  The magnetic field
in the target cell was directed along the $y$ axis, and its magnitude ranged
from 70 mT in the center to 40 mT at the ends.

A more complete description of the laser-driven polarized deuterium source
used in this experiment can be found elsewhere \cite{bibl:ldt}.  The main
points are summarized here.  Deuterium molecules were dissociated and
the resulting atoms were mixed with a potassium vapor in a spin-exchange
cell. The valence electrons of the
potassium atoms were polarized by optical pumping with
circularly-polarized light from a titanium-sapphire laser.  The target
polarization was reversed at intervals of about 30 seconds
by switching the helicity of the laser light
and slightly retuning the laser to the appropriate frequency.
A 70~mT
magnetic field in the cell inhibited radiation trapping in the potassium
vapor.  The magnetic field was parallel to the
$y$ axis and defined the
polarization axis for the source.  The deuterium electrons were polarized
by collisions between deuterium and potassium atoms in which the spins of
the valence electrons in the two atoms are exchanged.
This spin-exchange mechanism also polarized the
deuterium nuclei through deuterium-deuterium \mbox{(D-D)} collisions as a
result of the
small mixing between the electron and nuclear spins still present at 70~mT
\cite{TW-LWA:ste}.  The \mbox{D-D} spin-exchange distributed angular
momentum
between the electrons and nuclei, bringing the system toward a state of
spin-temperature equilibrium.  Previous work verified that
spin-temperature equilibrium is reached in laser-driven hydrogen
\cite{erlangen-ste} and deuterium \cite{argonne-ste} sources. 

The most accurate measurement of the beam polarization was obtained by
using a previous measurement of \ayp\ in $\vec{p}d$ elastic scattering
at 198.6 MeV proton energy and at a
recoil deuteron lab angle of $42.6^\circ$ \cite{stephenson}.  The beam
polarization has a relative statistical error of 1.1\%,
and the previous
measurement had a total error of 0.4\%.  The different beam energies in
the two experiments contribute an error of 0.3\%.  Added in quadrature,
these errors give an overall relative error of 1.2\% in the normalization
of \ayp .
During the
experiment, the beam polarization varied between 0.61 and 0.75.

The target polarization was determined by measuring the polarization
asymmetries in the deuteron breakup reaction. 
In the plane-wave impulse approximation, the spin observables for
proton knockout are equal to the well-known spin observables 
for \vecpp\ elastic
scattering at the center of mass energy of the two protons.  A Monte Carlo
calculation using a deuteron momentum wave function derived from the
Argonne V18 potential \cite{forest-priv} 
was used to determine the correction to the spin observables due to the 
momentum of the proton within the deuteron.  The momentum of the 
outgoing neutron was restricted to 
\mbox{\( | \textbf{p}_n | < 60 \) MeV/c}
so that the deuteron $D$ state contribution could be ignored.  The 
results were not sensitive to the maximum neutron momentum.  However, 
the beam polarization measured using this technique was approximately 
13\% below the value measured in the elastic reaction as 
described above.  For this reason the relative
systematic error in the target polarization was assigned to be 13\%.

The uncertainty in the target polarization is the dominant systematic
error in \itll\ and \cyy .  Systematic 
errors in the angle
reconstruction are less than 10 mr, and this leads to the largest 
systematic error in \ayp\ of $\pm 0.007$.  Errors
due to particle identification, background subtraction, luminosity
normalization, and beam energy have also been considered.  The total
systematic error from all sources other than the polarizations is
$\pm 0.010$ for \ayp , $\pm 0.011$ for \itll , and $\pm 0.015$ for
\cyy .

\begin{figure}[bt]
\centering{\epsfig{file=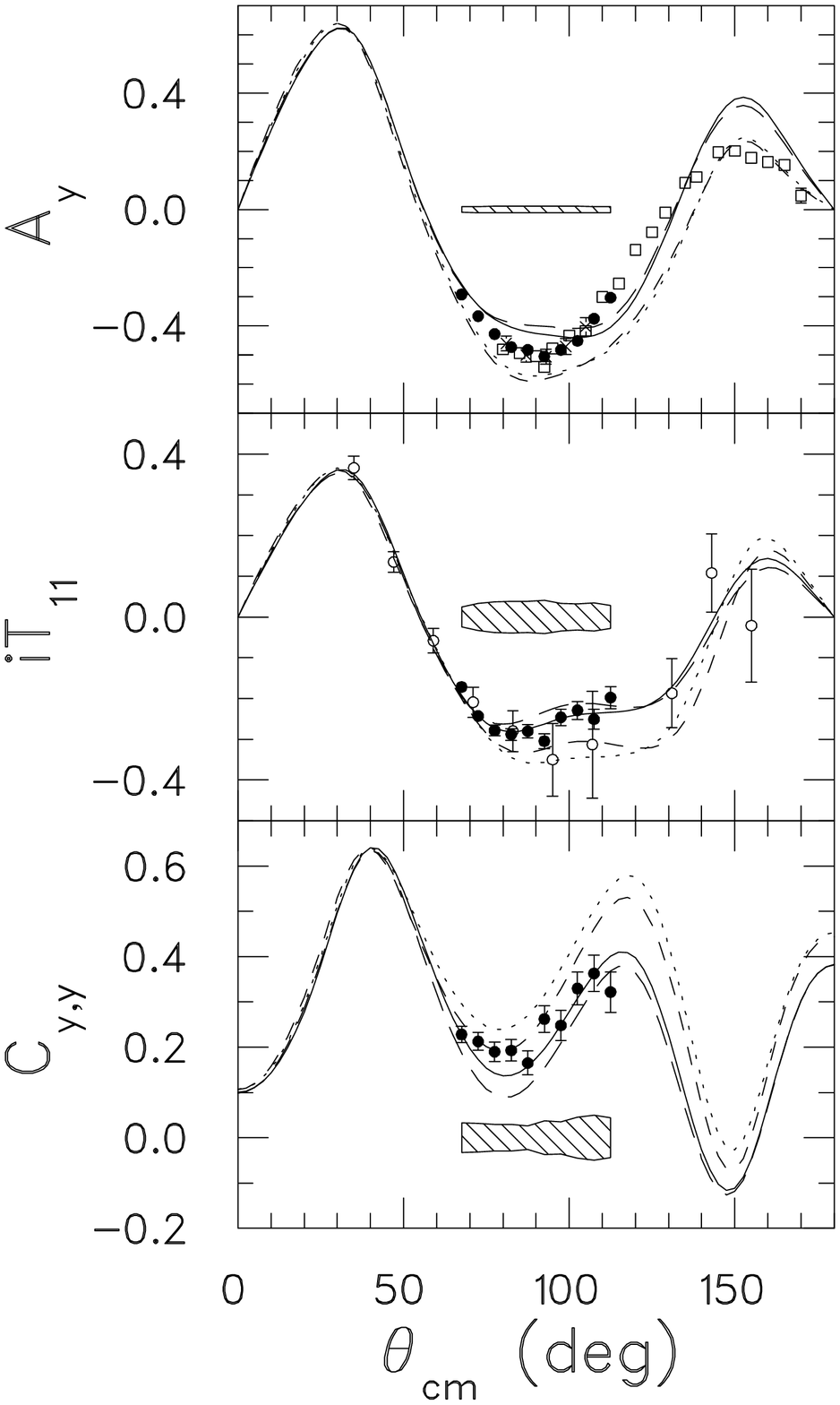,width=3.3in}}
\caption{Spin observables in \vecpd\ elastic scattering at a proton lab
energy of 197 MeV as a function of
the proton scattering angle in the center of mass reference frame,
$\theta_{\mathrm{cm}}$. The error bars indicate statistical errors and the
systematic error is indicated by the hashed area.  The 
top panel shows the the proton analyzing power \ayp ,
the center panel
shows the deuteron vector analyzing power \itll , and
the bottom panel shows
the vector-vector spin correlation parameter \cyy .
The results of this work are indicated by solid circles.  Previous
measurements of \ayp\ from Rochester (squares) \protect\cite{ab72}
and IUCF (crosses) \protect\cite{iucf93} are also shown, as well
as the measurement of \itll\ with a 395 MeV deuteron beam scattering
from a proton
target at Saclay (open circles) \protect\cite{garcon}.  The
theoretical curves are calculations using CD-Bonn+TM (solid), AV18+TM
(long dashed), CD-Bonn only (dotted), and AV18 only (short dashed).}
\label{fig:results}
\end{figure}

The spin observables measured in this experiment have been compared to
predictions both with and without the addition of a 3NF. These predictions
require a calculation of the transition amplitude for elastic
nucleon-deuteron scattering, which includes the nucleon exchange
term, the direct action of a 3NF, and rescattering interactions of three
nucleons through two- and three-nucleon forces. Details of the
computational methods and performance have been published
\cite{bochum-stuff}. The CD-Bonn \cite{Bonn-orig} and Argonne V18
\cite{av18-orig} potentials were both used.  Of the modern
two-nucleon potentials, these two are believed to be the most 
different \cite{pp-de+nu}.  The 3NF was
chosen to be the two-pion-exchange Tucson-Melbourne model.  In the 3NF,
the strong cutoff parameter $\Lambda$ is adjusted separately for each
two-nucleon potential to match the experimental triton binding energy
\cite{Glockle:triton}.  The numerical error in solving the
Schr\"{o}dinger equation for $Nd$ scattering is less than 2\%.  The
calculations were performed at an incident nucleon energy of 190~MeV.  
The resulting predictions
are shown along with the experimental results in Figure \ref{fig:results}.

The top panel of Figure \ref{fig:results} shows the proton analyzing
power \ayp .  These results are consistent with the previous measurements,
and fall in between the two predictions.
The center panel of Figure \ref{fig:results} shows the deuteron analyzing
power $iT_{11}$.  
The pure two-nucleon force calculations disagree with these
results, but the calculations including the 3NF are consistent with the
results.  The bottom panel of Figure \ref{fig:results} shows the 
spin correlation parameter $C_{y,y}$.  
Again, the 2NF calculations disagree with the results at the largest
angles, but the Argonne V18 calculation without the 3NF is consistent with
the data at the smallest angles.  

In conclusion, polarization observables in \vecpd\ elastic scattering have
been measured at a proton energy of 197~MeV over the angular range
\( 65^\circ < \thcm < 115^\circ \). This was the first experiment
to use a laser-polarized deuterium target. 
Calculations with the two-nucleon force alone do not reproduce the data,
but the inclusion of the Tucson-Melbourne 3NF improves the agreement
with data.
These results provide evidence for the nuclear three-body force;
however, further theoretical and experimental work is needed to fully
characterize its spin dependence. 

We thank J. L. Friar and V. R. Pandharipande for useful comments.  We
thank the staff and operators at IUCF, especially G. East, T. Sloan, R.
Pollock, and J.~Doskow, for their help and for the high quality of the
beam we received.  We also thank R. S. Kowalczyk, \mbox{Z.-T. Lu,} 
K. Bailey, G.~Smith, and
A.~Kenyon for assisting with the development of the laser-driven source,
D. Tupa for the loan of the two titanium-sapphire lasers, C.~A.~Miller for
the loan of the four wire chambers, and our glass blower, W. Lawrence.
This work was supported by the U.~S. National Science Foundation, the
U.~S. Department of Energy, the German Bundesministerium f\"ur
Bildung, Wissenschaft, Forschung und Technologie, and the Polish
Committee for Scientific Research.  The calculations were
performed on the CRAY T90 and T3E at the John von Neumann Institute for
Computing, J\"ulich, Germany.

% now the references. delete or change fake bibitem. delete next three
%   lines and directly read in your .bbl file if you use bibtex.

\end{document}